\documentclass[sn-nature]{sn-jnl}

\usepackage{graphicx}%
\usepackage{multirow}%
\usepackage{amsmath,amssymb,amsfonts}%
\usepackage{amsthm}%
\usepackage{mathrsfs}%
\usepackage[title]{appendix}%
\usepackage{xcolor}%
\usepackage{textcomp}%
\usepackage{manyfoot}%
\usepackage{booktabs}%
\usepackage{algorithm}%
\usepackage{algorithmicx}%
\usepackage{algpseudocode}%
\usepackage{listings}%
\usepackage{tabularx}

\theoremstyle{thmstyleone}%
\theoremstyle{thmstyletwo}%

\theoremstyle{thmstylethree}%

\begin{document}

\title[Recent Progress on Interstellar Radionuclides on Earth and the Moon]{Recent Progress on Interstellar Radionuclides on Earth and the Moon}


\author*[1,2,3]{\fnm{Dominik} \sur{Koll}}\email{d.koll@hzdr.de}
\author[2]{\fnm{Sebastian} \sur{Fichter}}
\author[4]{\fnm{Michael} \sur{Hotchkis}}
\author[5]{\fnm{Martin} \sur{Martschini}}
\author[5]{\fnm{Silke} \sur{Merchel}}
\author[1]{\fnm{Stefan} \sur{Pavetich}}
\author[2,6]{\fnm{Annabel} \sur{Rolofs}}
\author[1]{\fnm{Steve} \sur{Tims}}
\author[2,3]{\fnm{Sebastian} \sur{Zwickel}}
\author[2,3]{\fnm{Anton} \sur{Wallner}}

\affil[1]{\orgdiv{Department of Nuclear Physics and Accelerator Applications}, \orgname{\newline The Australian National University}, \orgaddress{\city{Canberra}, \postcode{2601}, \country{Australia}}}

\affil[2]{\orgdiv{Accelerator Mass Spectrometry and Isotope Research}, \orgname{Helmholtz-Zentrum Dresden-Rossendorf}, \orgaddress{\city{Dresden}, \postcode{01328}, \country{Germany}}}

\affil[3]{\orgdiv{Institute of Nuclear and Particle Physics}, \orgname{TUD Dresden University of Technology}, \orgaddress{\city{Dresden}, \postcode{01069}, \country{Germany}}}

\affil[4]{\orgdiv{Centre for Accelerator Science}, \orgname{Australian Nuclear Science and Technology Organisation}, \orgaddress{\city{Sydney}, \postcode{2232}, \country{Australia}}}

\affil[5]{\orgdiv{University of Vienna - Faculty of Physics, Isotope Physics}, \orgaddress{\city{Vienna}, \postcode{1090}, \country{Austria}}}

\affil[6]{\orgdiv{Argelander-Institut für Astronomie}, \orgname{University of Bonn}, \orgaddress{\city{\newline Bonn}, \postcode{53121}, \country{Germany}}}


\abstract{The detection of interstellar radionuclides in geological archives provides insights into nucleosynthesis in stars and stellar explosions as well as interstellar medium dynamics in the Local Bubble and the Local Interstellar Cloud. In this work, current projects to detect interstellar radionuclides with accelerator mass spectrometry will be reviewed. These projects aim to address unsolved questions regarding the timing and the origin of the influxes and to establish new radionuclides for future searches. For the first time, experimental evidence for an inhomogeneous deposition of interstellar $^{60}$Fe on Earth will be presented and another potential source for $^{60}$Fe on Earth and the Moon, primary galactic cosmic rays, will be introduced.}




\maketitle

\section{Introduction}\label{intro}

\begin{figure}[H]
\centering
\includegraphics[width=0.95\textwidth, trim={0cm 0cm 0cm 0cm}, clip]{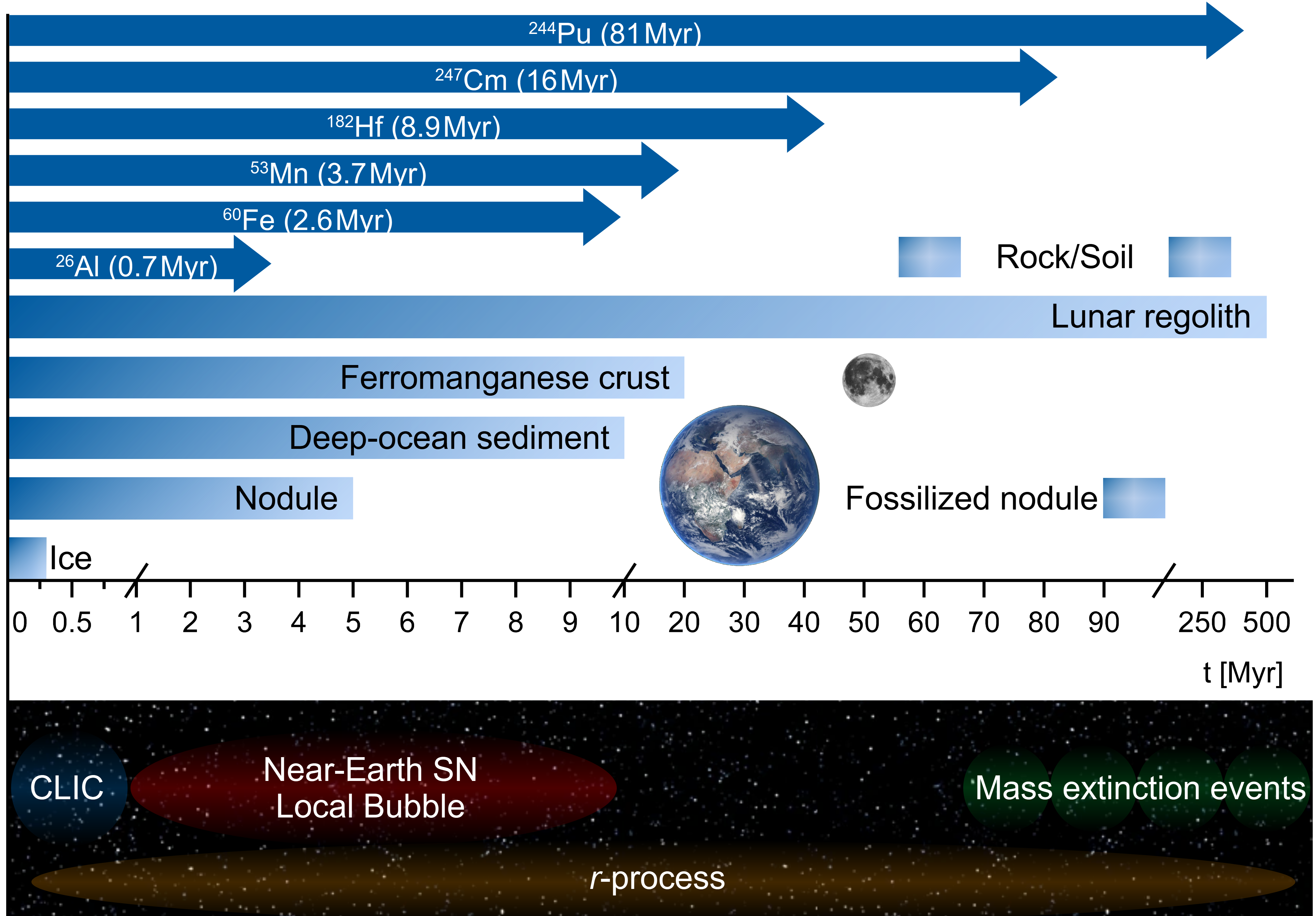}
\caption[]{Schematic overview of interstellar radionuclides with their half-lives, potential detection timescales, geological samples and interstellar structures such as the complex of local interstellar clouds (CLIC) and the Local Bubble being investigated in past, current and future projects. }
\label{fig:Summary}
\end{figure}

Structure formation and chemical evolution in the universe are driven by stellar nucleosynthesis and explosions. Freshly synthesised radioactive isotopes provide important information about these processes either through their decay in the interstellar medium and subsequent detection of their decay radiation, or through their live presence as ions in cosmic rays or as condensed dust.\\
The radionuclide $^{60}$Fe (t$_{1/2}$\,=\,2.6\,Myr \cite{Rugel2009, Wallner2015}) is arguably the most successfully investigated live interstellar radionuclide near Earth. Its presence was shown through an elevated $^{60}$Fe/Fe ratio by single-atom counting (i.e.\ accelerator mass spectrometry (AMS), \cite{Kutschera2023, Wallner2023, Fields2023}) in several deep-ocean samples such as sediments \cite{Wallner2016, Ludwig2016}, ferromanganese crusts and nodules \cite{Knie2004, Wallner2016, Wallner2021}, in Antarctic surface snow \cite{Koll2019a, KollMSc} and even on the Moon \cite{Fimiani2016} (Figure \ref{fig:Summary}). Furthermore, live $^{60}$Fe nuclei were detected in galactic cosmic rays (GCR) near Earth \cite{Binns2016} as well as $\gamma$-rays from $^{60}$Fe decay \cite{Wang2020}. All investigations combined yield a clear picture of ongoing nucleosynthesis in the Milky Way and the solar system's encounter with an $^{60}$Fe-filled interstellar medium. Massive stars synthesize $^{60}$Fe in their advanced burning stages through the weak \textit{s}-process \cite{Limongi2006a, Woosley2007}. When massive stars undergo a core-collapse supernova, the synthesized $^{60}$Fe is ejected into interstellar space. The hypothesis of a transient passage of supernova ejecta over the solar system was disfavoured after the duration of the $^{60}$Fe influx was shown to be on the order of Myr \cite{Fitoussi2008, Wallner2016}. A transient passage is still possible, however, requires a more complicated treatment of the interaction of the supernova ejecta with the interstellar medium. A condensation into dust grains was established as a model to overcome the heliospheric shielding of the solar system \cite{Athanassiadou2011, Fry2016}. The proposed regions hosting the supernova(e) responsible for the deposited $^{60}$Fe on Earth include Scorpius-Centaurus and the Tucana-Horologium group, however, the search for a runaway star from these supernovae was so far inconclusive \cite{Neuhaeuser2012, Neuhaeuser2020}. It was argued that several supernovae in the Local Bubble contribute to the measured $^{60}$Fe on Earth and that the deposition of interstellar radionuclides is linked to the dynamics of the interstellar medium within and adjacent to the Local Bubble \cite{Schulreich2017, Schulreich2023}. \\
A second interstellar radionuclide, $^{244}$Pu (t$_{1/2}$\,=\,81\,Myr \cite{Aggarwal2006}), was only recently discovered on Earth in a ferromanganese crust together with $^{60}$Fe using AMS  \cite{Wallner2021}. The comparison of both radionuclides could shed light onto the production site of the heaviest elements in the universe, the so-called \textit{r}-process. Furthermore, the influx profiles of both radionuclides should reflect interstellar medium dynamics and mixing of ejecta from different types of stellar explosions \cite{Wehmeyer2023}.\\
Cosmogenic $^{53}$Mn is often used as a proxy for cosmogenically produced $^{60}$Fe and to quantify the surplus of interstellar $^{60}$Fe \cite{Koll2022b, Fimiani2016}. Both radionuclides are measured at the Heavy-Ion Accelerator Facility (HIAF) of the Australian National University (ANU) \cite{Wallner2023}, which is currently the only facility capable of measuring these radionuclides with the required sensitivity after the closure of the Munich accelerator lab \cite{Koll2019a}. The measurements of $^{244}$Pu require uttermost sensitivity and, therefore, the dedicated setup of VEGA at the Australian Nuclear Science and Technology Organisation (ANSTO) \cite{Hotchkis2019} with the highest sensitivity for actinide measurements \cite{Koll2022} is currently used. In the future, actinide measurements are expected to be continued at the newly installed Helmholtz Accelerator Mass Spectrometer Tracing Environmental Radionuclides (HAMSTER) facility at Helmholtz-Zentrum Dresden-Rossendorf (HZDR), where the measurement of further radionuclides such as $^{182}$Hf might also become feasible. Besides interstellar radionuclides, the Dresden AMS (DREAMS) facility at HZDR is currently being used to measure cosmogenic radionuclides such as $^{10}$Be, $^{26}$Al and $^{41}$Ca \cite{Rugel2016,Lachner2023,Vilches2023} from the same samples. These radionuclides are used to characterize the samples with respect to accumulation of radionuclides and to date them.\\
The search for interstellar radionuclide depositions on Earth has proven to be successful, though with intrinsic limitations. First, the low influx of interstellar radionuclides leads to ultra-low concentrations and therefore low counting statistics of single atoms when using AMS. The resulting larger uncertainties give room for interpretations which can only be narrowed down by analysing more samples to reduce statistical fluctuations. Second, the differences between geological archives could lead to variations in the interstellar radionuclide concentrations. Again, more samples from different locations are required to constrain these effects. Third, the calculation of an interstellar radionuclide fluence based on radionuclide concentrations in geological archives depends on several estimations. The influx, distribution and incorporation of interstellar radionuclides need to be quantified. So far, a globally homogeneous deposition was assumed with the only difference between archives being the incorporation efficiency for interstellar radionuclides. In this work we will show that a globally inhomogeneous deposition of radionuclides needs to be considered.
\\
In the following, the current status of research, summarising previous results and highlighting open questions, will be discussed and recent projects will be presented briefly with their future impact on our understanding of interstellar radionuclide influxes.

\newpage

\section{Interstellar radionuclides on Earth and the Moon}\label{sec1}

\subsection{Million-year-old interstellar influxes} \label{crust}
Deep-ocean samples were shown to be ideal archives for interstellar radionuclides over timescales of several million years. The initial discovery of an interstellar $^{60}$Fe influx 2\,--\,3\,Myr ago into deep-ocean ferromanganese crusts \cite{Knie1999a, Knie2004} sparked numerous investigations on different deep-ocean archives and different interstellar radionuclides over the past 25\,yr. Most notably, this $^{60}$Fe influx was discovered on a global scale in ferromanganese crusts, nodules and deep-ocean sediments \cite{Wallner2016} and a second influx around 7\,Myr ago was recently found \cite{Wallner2016, Wallner2021}. While the 2\,--\,3\,Myr influx was detected numerous times and the timing could be constrained, the second influx was so far detected only once peaking at 6.3\,Myr \cite{Wallner2021}, while another indication for the second influx was found in the time interval 6.5\,--\,8.7\,Myr \cite{Wallner2016}. This discrepancy needs to be resolved by more accurately characterized and dated samples. \\
Recently, the discovery of a second interstellar radionuclide, $^{244}$Pu, was announced \cite{Wallner2021}. Both, $^{60}$Fe and $^{244}$Pu, were detected in the same ferromanganese crust sample. The time-resolution of the $^{244}$Pu data, however, only represents an integral over the more resolved $^{60}$Fe data and a higher time-resolution would be required to see if the $^{244}$Pu influx follows the time profile of the $^{60}$Fe influx.\\
A statistical indication for a third interstellar radionuclide, $^{53}$Mn, was reported in a compilation of four different ferromanganese crust data sets \cite{Korschinek2020}. This indication requires a more detailed analysis with higher statistics and well-characterized samples to confirm or refute the presence of interstellar $^{53}$Mn on Earth.\\
The search for interstellar $^{26}$Al in a deep-ocean sediment core concomitant with the influx of $^{60}$Fe was so far unsuccessful \cite{Feige2018} and should be revisited in the future with an improved $^{26}$Al detection to discriminate interstellar from cosmogenic $^{26}$Al. \\\\
Recently, a large 3.7\,kg deep-ocean ferromanganese crust sample was acquired to address these open questions \cite{KollPhD}. The characterization of the crust focused on establishing an accurate age model to constrain the timing of the second $^{60}$Fe influx as well as to compare the influx profiles of $^{60}$Fe and $^{244}$Pu. Three drill-holes were taken with a depth resolution of 1\,--\,2\,mm corresponding to a time resolution of $\pm$\,0.25\,Myr per sample up to 10\,Myr for $^{10}$Be dating and $^{60}$Fe measurements \cite{Koll2022}. The remaining crust sample was split into layers of $\pm$\,0.5\,Myr time resolution for the measurement of $^{244}$Pu with comparatively lower abundance than $^{60}$Fe. This investigation will be complemented by $^{26}$Al, $^{53}$Mn, $^{182}$Hf and $^{247}$Cm measurements in the future, with $^{247}$Cm analysis almost finished. The results and their interpretation will be subject of a subsequent publication.

\subsection{Recent interstellar influxes} \label{ice}
The previously described deep-ocean samples with their intrinsically low time resolution on the order of kyr to Myr are less suitable to investigate $^{60}$Fe influxes at recent times. A detection of $^{60}$Fe on the yr to kyr timescale requires higher sample masses due to shorter accumulation times. These larger samples, however, need to be concentrated without diluting $^{60}$Fe with terrestrial Fe beyond detectability. Polar snow and ice, in particular from Antarctica, almost exclusively meet these requirements with stable Fe concentrations of only ng/g \cite{Liu2024}. A present-day $^{60}$Fe influx was discovered in a total of 500\,kg of Antarctic surface snow \cite{KollMSc, Koll2019a}. This influx was confirmed by deep-ocean sediment surface samples on a timescale up to 33\,kyr \cite{Wallner2020}. The origin of this recent $^{60}$Fe influx was discussed in terms of either a fade-out of the 2\,--\,3\,Myr old influx, a reflection of the 2\,--\,3\,Myr old influx or an independent new reservoir of $^{60}$Fe in the form of the Local Interstellar Cloud (LIC) \cite{Wallner2020, Koll2019a, Koll2020}. It was proposed to search for $^{60}$Fe in "deep-sea sediments, fast-growing ferromanganese crusts as well as Antarctic and Arctic ice with focus on the time period between 50\,kyr and 300\,kyr" \cite{Koll2020}, because the solar system entered the LIC about 40\,kyr ago. \\\\
In order to reduce any terrestrial influences, a large Antarctic ice sample from the European Project for Ice Coring in Antarctica (EPICA) EDML ice core, adjacently located to the Antarctic snow sample, was acquired. A total of about 300\,kg of 40\,--\,80\,kyr old Antarctic ice from continuous flow analysis was chemically processed \cite{Fichter2024}. After evaporation and stable element analysis by inductively-coupled plasma mass spectrometry (ICP-MS), purified elemental fractions were prepared for the measurement of atmospheric $^{10}$Be and $^{26}$Al, interplanetary $^{41}$Ca and $^{53}$Mn, as well as interstellar $^{60}$Fe. At this time, all measurements are completed and data evaluation and interpretation are in progress \cite{RolofsMSc}. The abundance of interplanetary $^{53}$Mn in addition to cosmogenic $^{10}$Be might also shed light on the total mass of deposited interplanetary dust onto Earth \cite{Rodrigues2018}.
In the future, a set of ice core samples and in particular deep-ocean sediments should be used to extend the data set to the time period 100\,kyr to 300\,kyr, where time resolution becomes less critical and sediments become more easily accessible than ice cores.

\subsection{Interstellar influxes onto the Moon} \label{lunar}
Lunar soil presents a unique archive for interstellar radionuclides, because, unlike on Earth, the lunar surface is not subject to permanent alterations through geological activity. Therefore, any interstellar dust arriving on the Moon is implanted into its surface and accumulates over long time-scales. Integration times up to a few hundred million years can be reached \cite{Fields2023}. However, continuous (micro-)meteoritic bombardment causes a mixing of the soil to higher depths in the range of cm for a mixing timescale of Myr. This process is called the gardening-effect \cite{Nishiizumi1979} and lunar gardening needs to be taken into account for the calculation of radionuclide fluences. Additionally, cosmic-rays impinging onto the lunar surface cause nuclear reactions and thereby produce cosmogenic radionuclides such as $^{10}$Be, $^{26}$Al, $^{41}$Ca, $^{53}$Mn, as well as $^{60}$Fe \cite{Li2017, Reedy1972, Fimiani2016}. To date, there has been only one successful detection of interstellar radionuclides in lunar soil by AMS. Fimiani et al. \cite{Fimiani2016} found an enhanced concentration of interstellar $^{60}$Fe above a cosmogenic $^{60}$Fe base-level.\\\\
We started a project to extend the search for interstellar radionuclides in lunar soil with a particular focus on interstellar $^{244}$Pu and $^{247}$Cm. Being pure \textit{r}-process nuclides, they have the potential to set new constraints on both the frequency and production yields of recent \textit{r}-process events in the solar neighbourhood. First results within this project of cosmogenic radionuclides in lunar soil from the Apollo missions were recently presented \cite{Zwickel2024}.

\section{Extraction and purification of interstellar radionuclides}

Interstellar radionuclides are deposited on Earth and the Moon in minuscule quantities. Stable elements as well as cosmogenic and anthropogenic radionuclides typically obscure any interstellar radionuclide influx signal by many orders of magnitude in intensity. The AMS measurements, except for actinides, rely on the measurement of the radionuclides to a stable reference isotope which means that a large stable element abundance leads to low isotopic ratios and, therefore, in some cases non-detectability. It was shown above that the choice of a suitable sample material alleviates the situation. However, it is still necessary to chemically extract the minuscule quantities from the sample matrix.\\\\
The chemical extraction of long-lived radionuclides from environmental samples is well-developed for cosmogenic radionuclides such as $^{10}$Be or $^{26}$Al. Procedures for samples such as meteorites \cite{Merchel1999}, lunar soil \cite{Nishiizumi1984}, deep-ocean ferromanganese crusts \cite{Koll2022} or sediments \cite{Bourles1989} are typically used as a starting point to develop dedicated extraction procedures for interstellar radionuclides. The advances of chemical preparation techniques for interstellar radionuclides are illustrated through the examples $^{60}$Fe, $^{244}$Pu, $^{247}$Cm, and $^{182}$Hf.\\
The earliest studies of interstellar radionuclides on Earth by AMS already focused on $^{60}$Fe and $^{244}$Pu \cite{Knie1999a, Knie2004, Wallner2004, Paul2001}. The AMS measurement of $^{60}$Fe suffers from its stable isobar $^{60}$Ni, which is typically more than 10 orders of magnitude more abundant in environmental samples than $^{60}$Fe. The chemical extraction of $^{60}$Fe is tailored to extract $^{60}$Fe quantitatively, while suppressing Ni from the matrix as much as possible. This is achieved through selective hydroxide precipitations with ammonia, where Ni forms a soluble ammine-complex and is efficiently separated from Fe. Subsequently, Fe is further purified by the formation of a chloro-complex and its binding to an anion-exchange resin, while residual Ni is non-binding. This basic chemistry was refined and adapted to a large variety of different environmental matrices \cite{Merchel1999, Koll2022, Fichter2024, KollMSc, RolofsMSc}. Further suppression of Ni is required in the AMS measurement, where the use of a gas-filled magnet has proven to be most effective \cite{Wallner2023}. \\ Isobaric suppression is not necessary when extracting $^{244}$Pu, because trans-uranium elements do not have stable isobars. There is also no dilution by stable elements due to the trace-abundance of Pu in the environment. The challenge for chemistry is reduced to achieving highest recoveries from largest matrices. The benevolent chemical property of Pu to exhibit several adjustable oxidation states makes a separation from ubiquitous matrix elements possible. Matrix reduction steps include precipitations as well as solvent extractions. Eventually, the oxidation state of Pu is adjusted to +IV, which leads to a strong binding to an anion-exchange resin in chloric and nitric media. Virtually all other matrix elements are separated from Pu and subsequently Pu is recovered by selective reduction to the non-binding +III oxidation state. This highly efficient chemical procedure is applicable to large samples, even kg sized ferromanganese crusts \cite{Knie2004, Wallner2021, Koll2022}.\\\\
The conditions to extract the radionuclide $^{247}$Cm are similar to $^{244}$Pu, i.e.\ it has no stable isobar, there is no dilution and the only requirement for chemistry is to obtain highest recovery of Cm after matrix separation. However, the chemical properties of Cm have so far made this seemingly easy task challenging. Curium shows a similar chemical behaviour to Am, which is often used as chemical yield tracer and internal reference \cite{Christl2014}. Both trans-uranium elements possess the +III oxidation state and are, hence, not retained on the anion exchange resin used for Pu separation, similar to the other matrix elements. However, the use of a diglycolamide based resin successfully separates the trivalent actinides Am and Cm from the majority of the matrix elements \cite{Fichter2024}. Due to the high amount of lanthanides in some samples (e.g.\ 50\,g ferromanganese crust contain up to 100\,mg of lanthanides) another separation step is required in order to suppress lanthanides in the final Am and Cm fraction as much as possible. This also helps to minimize interferences in the final AMS measurement. Thus, a separation based on the different complex stability of trivalent lanthanide/actinide thiocyanate complexes is used \cite{Luisier2009, Fichter2024}. Currently, further tracer tests of Am and Cm separation, at the sensitivity level provided by AMS, are being conducted in order to assess potential differences in the extraction and measurement efficiencies and to warrant the use of $^{243}$Am as a suitable tracer for future $^{247}$Cm measurements. An isotopically pure $^{248}$Cm tracer is also currently being produced by element selective resonance ionization, isotope selective mass separation and subsequent implantation.\\\\
The radionuclide $^{182}$Hf is of particular interest for nuclear astrophysics as its production has \textit{s}- and \textit{r}-process components and its half-life is intermediate compared to $^{60}$Fe and $^{244}$Pu. The AMS measurement is challenging due to its abundant stable isobar $^{182}$W. An efficient isobaric background reduction was achieved using fluoride molecules and a helium/oxygen buffer-gas-filled radio-frequency quadrupole \cite{Martschini2020} with the newly developed AMS-based method Ion-Laser InterAction Mass Spectrometry (ILIAMS) \cite{Martschini2022}. Nevertheless, varying W concentrations from samples still pose unpredictable variations in the ultimate sensitivity of ILIAMS. The isobaric interference from $^{182}$W can arise from the ion-source and its metallic components, the metallic sample holders and from the processed sample itself. Unfortunately, the chemical reduction of W as well as the separation of Hf from the environmental matrix is more challenging than for the radionuclides presented above. 
This complexity arises from hafnium's inherent chemical properties being a high-field-strength transition metal. Thus, dissolving and maintaining Hf in acidic solutions remains an ongoing challenge which is currently being tackled by combined efforts at HZDR and at Vienna Environmental Research Accelerator (VERA).
Indeed, the acid leaching step, which is sufficient to solubilize the above mentioned radionuclides $^{60}$Fe, $^{244}$Pu and $^{247}$Cm from different sample matrices, does not maintain the majority of the Hf in solution. This could be shown by both, ICP-MS measurements of the resulting leachate and redissolved residue and through neutron activation analysis of the residue. Thus, selective leaching of Hf from the sample residue is the key to further separate it from remaining matrix elements. Currently, different separation strategies are being tested and optimized in order to subsequently suppress isobaric W while retaining a high yield of Hf.

\section{Interstellar radionuclides in cosmic rays}
Interstellar radionuclides are also present in galactic cosmic rays (GCR). Binns et al. \cite{Binns2016} analysed the isotopic composition of cosmic rays striking the NASA Advanced Composition Explorer - Cosmic Ray Isotope Spectrometer (ACE-CRIS) instrument for almost 17 years since 1997. A total of 15 $^{60}$Fe nuclei were identified with energies between 195\,MeV/nucleon and 500\,MeV/nucleon. This can be compared to the recorded 3.55\,$\times$10$^{5}$ Fe nuclei, leading to a ratio $^{60}$Fe/Fe ratio in GCR of 3.9\,$\times$10$^{-5}$ after corrections \cite{Binns2016} near Earth and 6.2\,$\times$10$^{-5}$ at the GCR acceleration source after accounting for radioactive decay and losses through leakage or nuclear reactions in the interstellar medium \cite{Binns2016}. A supernova in the Scorpius–Centaurus (Sco–Cen) OB association less than 500\,kyr ago could be the acceleration source for these $^{60}$Fe nuclei \cite{deSereville2024}. \\\\
An obvious question is: Does primary cosmic ray $^{60}$Fe account for or contribute to the $^{60}$Fe concentrations we find on Earth or the Moon? Note, here we consider primary $^{60}$Fe in cosmic rays, in contrast to the secondary cosmic ray produced $^{60}$Fe, i.e.\ cosmogenic $^{60}$Fe in meteorites, interplanetary dust or on the Moon \cite{Knie1999b, Koll2019a, Fimiani2016, Koll2022b}. 
Assuming a constant $^{60}$Fe/Fe ratio in GCR near Earth over millions of years of 4$\times$10$^{-5}$, one can estimate the flux of $^{60}$Fe nuclei. The Fe/H ratio in GCR is about 2$\times$10$^{-4}$ \cite{Mewaldt1994}, about one order of magnitude higher than the solar abundances \cite{Lodders2020}. The $^{60}$Fe/H ratio in GCR near Earth becomes 8$\times$10$^{-9}$. To obtain the flux of $^{60}$Fe nuclei, this ratio needs to be scaled by the total flux of GCR which mainly consists of protons. The flux of protons depends on the considered energy of the cosmic rays and is on the order of 2 protons/cm$^{2}$/s \cite{McKinney2006, Crites2013, Naito2020} for the energy range up to 1\,GeV/nuc. The flux of $^{60}$Fe in GCR at 1\,AU is therefore roughly 0.5 $^{60}$Fe nuclei/cm$^{2}$/yr in the energy range 195\,--\,500\,MeV/nucleon. \\\\
This value can now be compared with the determined $^{60}$Fe depositions in various geological and lunar archives. However, highly energetic heavy ions in GCR may undergo nuclear reactions in any dense medium. We used a simulation of cosmic ray propagation and particle production in Earth's atmosphere with the tool DYASTIMA \cite{Paschalis2014} to qualitatively verify that there is a surviving primary $^{60}$Fe cosmic ray component. Simulating 1\,GeV/nuc Fe ions impinging onto Earth's atmosphere showed charge-changing as well as mass-changing reactions into lighter elements indicated by the appearance of Mn, Cr, V, Ti, and lighter Fe isotopes, respectively, in the spectra. Simulating 100\,MeV/nuc though showed a strongly reduced signature of nuclear reactions resulting in Fe ions with energies well below 10\,MeV/nuc, indicating stopping and survival. This is consistent with a sharp drop of the mass and charge-changing cross-section below 10\,MeV/nuc for Fe ions \cite{Luoni2021}. To obtain a more quantitative survival rate, the range of Fe ions with a given energy in air can be calculated using SRIM \cite{Ziegler2010} and the survival rate of the ions is approximated by $e^{-\sigma\rho}$ with $\sigma$ being the mass and charge-changing cross-section and $\rho$ the target density. Assuming a constant cross-section of 1.8\,b \cite{Luoni2021}, the energy-dependent survival rate is displayed in Table \ref{tab:Survival}. Further on, the survival rate needs to be scaled with the differential flux of Fe ions, estimated from simulated and measured galactic cosmic ray spectra \cite{Lyu2024, Keating2012}. The energy range up to 1\,GeV/nuc accounts for only about 10$\%$ of the total GCR Fe flux, however, more energetic ions will mostly not survive. The resulting total survival fraction for the energy range up to 1\,GeV/nuc is displayed in Table \ref{tab:Survival} which sums up to 35$\%$ of incoming GCR $^{60}$Fe with energies up to 1\,GeV/nuc. For the full spectrum, this would reduce to about 3.5$\%$ considering no survival for energies above 1\,GeV/nuc.\\

\newcolumntype{Y}{>{\centering\arraybackslash}X}
\begin{table*}
\centering
\begin{tabularx}{\textwidth}{YYYYY}  
\hline    \hline \\[-2ex]   
\textbf{Incident energy} & \textbf{Stopping density} & \textbf{Survival rate} &  \textbf{Flux fraction} & \textbf{Total survival} \\
 MeV/nuc & 10$^{23}$ at/cm$^{2}$ &  $\%$ & $\%$ & $\%$ \\
\hline \\[-2.5ex]
100 & 0.3 &  94 & 3.2 & 3.0  \\
200 & 1.1 & 81 & 6.0 & 4.9  \\
300 & 2.1 & 68& 8.6 & 5.8 \\
400 & 3.4 & 54& 10.3 & 5.6  \\
500 & 4.9 & 42& 10.9 & 4.6 \\
600 & 6.4 & 32& 11.6 & 3.7 \\
700 & 8.2 & 23& 12.0 & 2.8 \\
800 & 9.8 & 17& 12.4 & 2.1 \\
900 & 12 & 12& 12.4 & 1.5 \\
1000 & 14 & 9& 12.4 & 1.1 \\
\hline
0\,--\,1000 &  & & & 35 \\
\hline    \hline
\end{tabularx}
\caption{Energy dependent survival rate for GCR Fe nuclei impinging onto Earth's atmosphere. The stopping density was calculated by SRIM \cite{Ziegler2010} and the survival rate was estimated using a constant mass and charge-changing cross-section of 1.8\,barn \cite{Luoni2021}. The survival rate for each energy bin is folded with the flux fraction of the GCR spectrum up to 1000\,MeV/nuc to yield a total survival of GCR Fe ions in this energy regime. Considering the full GCR spectrum, the survival fraction would reduce to about 3.5$\%$.}
\label{tab:Survival}
\end{table*}

\noindent The data from Binns et al. \cite{Binns2016} covered energies between 195\,MeV/nuc and 500\,MeV/nuc, where a survival rate of about 60$\%$ would be expected. The detection of actinide GCR in Earth's orbit by a satellite showed that fragmentation corrections on the order of 14$\%$ were required \cite{Donnelly2012}, i.e.\ a survival rate of 86$\%$ when passing a detector setup with a total Lexan (C$_{16}$H$_{14}$O$_3$) equivalent surface density of 4.1\,g/cm$^{2}$. Hence, a significant survival fraction for primary galactic cosmic rays in Earth's atmosphere seems plausible.\\
This all shows that an energy-dependent primary GCR $^{60}$Fe fraction might indeed survive the stopping in the atmosphere or on the lunar surface and a more detailed future analysis of survival rates is required. In the course of this, the fragmentation of trans-iron cosmic rays into $^{60}$Fe should also be investigated and the cosmogenic production of $^{60}$Fe on trace atmospheric gases such as Kr should be revisited for completeness. \\\\
In this work, we compare the contribution of GCR $^{60}$Fe to geological $^{60}$Fe signals assuming 100$\%$ survival and deposition of GCR $^{60}$Fe. This conservative estimation of full survival yields an upper limit of the galactic cosmic ray contribution to $^{60}$Fe abundances in geological archives. The compiled fluences of $^{60}$Fe for all previously investigated deep-ocean samples are on the order of a few tens of $^{60}$Fe nuclei/cm$^{2}$/yr \cite{Wallner2016, Wallner2021, KollPhD}. An upper limit of GCR $^{60}$Fe deposition in deep-ocean samples can be calculated by taking the measured background rate of $^{60}$Fe in a ferromanganese crust of $<$\,0.07 $^{60}$Fe nuclei/cm$^{2}$/yr \cite{Wallner2021}, correcting it for an incorporation efficiency of 17$\%$ \cite{Wallner2021} and assuming a globally homogeneous deposition. The background upper limit of 0.4 $^{60}$Fe nuclei/cm$^{2}$/yr in the ferromanganese crust would be in agreement with the proposed amount of GCR $^{60}$Fe assuming full survival.\\
In this estimation, GCR $^{60}$Fe will not significantly contribute to the elevated $^{60}$Fe signals in deep-ocean samples. This is largely a result of the much larger $^{60}$Fe peak influxes and the time resolution of the archives. A contribution of GCR $^{60}$Fe could be realised by either analysing a very low influx of $^{60}$Fe, i.e.\ the recent $^{60}$Fe influx in Antarctica, or when integration times become much longer, i.e.\ on the Moon. The measured $^{60}$Fe deposition in Antarctica is 1.2 $^{60}$Fe nuclei/cm$^{2}$/yr \cite{Koll2019a}, higher but within the range of potential GCR $^{60}$Fe assuming full survival. The deposition of $^{60}$Fe on the Moon is 0.2\,--\,1\,$\times$10$^{8}$ $^{60}$Fe nuclei/cm$^{2}$ \cite{Fimiani2016} after decay-correction. The GCR contribution is time-dependent and accumulates over several Myr. The maximum secular equilibrium concentration for GCR $^{60}$Fe on the Moon after more than 10\,Myr is 1.9\,$\times$10$^{6}$ $^{60}$Fe nuclei/cm$^{2}$, less than the recorded deposition of $^{60}$Fe, but in some cases not negligible. The survival rate of GCR ions impinging onto the Moon should be slightly lower than onto Earth's atmosphere due to higher Z targets and consequently higher mass and charge-changing cross-sections. Long-lived GCR $^{244}$Pu can accumulate for several hundred million years until equilibrium is reached. This might be important to consider for future measurements of lunar soil.\\\\
In summary, GCR $^{60}$Fe is present outside of Earth's atmosphere and could potentially contribute to the $^{60}$Fe budget in geological and lunar archives due to their energy-dependent survival when stopping in matter. A more detailed investigation of the incident flux of GCR radionuclides as well as their energy-dependent survival rate when impinging onto the atmosphere or the lunar surface, however, is necessary to draw conclusive results. An anisotropic influx resulting from the deflection in Earth's magnetic field, analogous to the deflection of GCR that produce cosmogenic radionuclides in Earth's atmosphere with a latitudinal dependence, might become important in particular for the low-energy but surviving GCR $^{60}$Fe. Importantly, a variable primary $^{60}$Fe cosmic ray flux over timescales of millions of years should be considered. Upcoming investigations of Antarctic ice (see Section \ref{ice}) will yield further experimental limits on the potential deposition of GCR $^{60}$Fe and can be used to test future models of GCR $^{60}$Fe transport and fluences at different timescales. The results from Ludwig et al. \cite{Ludwig2016} from Pacific deep-ocean sediments represent the lowest $^{60}$Fe concentrations so far, however, do not represent the most stringent limit for GCR $^{60}$Fe due to an inhomogeneous deposition on a global scale.

\section{Inhomogeneous deposition of interstellar radionuclides}
The calculation of interstellar radionuclide fluences based on terrestrial deposition measurements relies on the knowledge of the incorporation efficiency, also known as uptake factor, of radionuclides into the respective geological archives as well as on the distribution factor for a globally inhomogeneous distribution. Atmospheric winds and ocean currents were proposed to be the main drivers of a potential inhomogeneous distribution of interstellar radionuclides on Earth \cite{Fry2016, Ertel2023}. Both factors require estimations. In previous work \cite{Knie2004, Fitoussi2008, Wallner2016, Ludwig2016, Wallner2021} a global inhomogeneity was not considered, i.e.\ an inhomogeneity factor of 1 was applied. In recent works, incorporation efficiencies into deep-sea crusts were calculated assuming 100$\%$ uptake in Indian Ocean sediments \cite{Wallner2016} and a globally homogeneous $^{60}$Fe distribution. Accordingly, any scaled ferromanganese crust $^{60}$Fe data becomes dependent on the Indian Ocean sediment's $^{60}$Fe fluence and would yield incorporation efficiencies on the order of 10$\%$\,--\,20$\%$ \cite{Wallner2016, Wallner2021, KollPhD}. \\\\
These assumptions were severely challenged after Pacific Ocean sediment cores \cite{Ludwig2016} showed a reduced $^{60}$Fe influx by 1\,--\,2 orders of magnitude compared to the Indian Ocean sediments and about a factor of 5\,--\,10 lower than into ferromanganese crusts without considering any incorporation efficiencies. To harmonize all measurements, it is either necessary to conclude that $^{60}$Fe was not quantitatively extracted from the Pacific Ocean sediments \cite{Ertel2024} or to accept that the global deposition of $^{60}$Fe is inhomogeneous \cite{Ludwig2016, KollPhD}. The chemical extraction of $^{60}$Fe from the Pacific Ocean sediment was based on a mild leaching, the citrate-bicarbonate-dithionite (CBD) technique \cite{LudwigPHD}. This leaching allows the specific dissolution and extraction of small-grained Fe particles, which was desirable for this specific project to reduce the dilution with stable terrestrial Fe as well as to extract the magnetosome fraction of the sediment, that was believed to contain enriched $^{60}$Fe after selective uptake of magnetotactic bacteria \cite{Bishop2011, Ludwig2016}. The stronger leaching of the Indian Ocean sediment on the other hand is commonly used to extract the authigenic fraction of sediments, i.e.\ the fraction that was previously dissolved in sea-water \cite{FeigePhD}. It needs to be shown if the difference in leaching is responsible for the measured difference in $^{60}$Fe influx.\\\\
In this work, a sample of the same Pacific sediment core used by Ludwig et al.\ \cite{LudwigPHD} was leached based on the stronger leaching procedure \cite{Bourles1989, FeigePhD} to investigate if $^{60}$Fe was previously lost as a result of the milder leaching procedure or if the $^{60}$Fe deposition is globally inhomogeneous. The 5.5\,g sediment sample from ODP 138 851 005H (37.0\,--\,37.5\,mbsf, 1.9\,Myr) was leached with the established procedure \cite{FeigePhD} for the Indian Ocean sediments and the dating was independently confirmed by cosmogenic $^{10}$Be at the DREAMS facility. After the leaching, the extracted authigenic Fe concentration in the sample was determined to be 0.8\,mg/g by ICP-MS. This is in contrast to the significantly reduced extraction of about 0.14\,mg/g with the milder CBD-technique \cite{LudwigPHD}. The Indian Ocean sediments had an extractable authigenic Fe concentration of about 2.4\,mg/g for the two sediment cores in the time window 1.7\,--\,2.0\,Myr. Assuming a homogeneous global deposition and a complete extraction, only the sediment characteristics, the sedimentation rate $g$, extractable Fe content $c$ and dry-mass density $\rho$ determine the final $^{60}$Fe/Fe ratio (Table \ref{tab:Sedimentdata}). The $^{60}$Fe/Fe ratio in the Indian Ocean sediment data serves as the reference. The expected ratio between the Pacific Ocean sediment and the Indian Ocean sediment for a globally homogeneous deposition can be calculated based on the sample requirement criterion introduced by Koll \cite{KollPhD}.
\begin{equation*}
    \frac{(^{60}Fe/Fe)_{Pacific}}{(^{60}Fe/Fe)_{Indian}} = \frac{(c \cdot \rho \cdot g)_{Indian}}{(c \cdot \rho \cdot g)_{Pacific}}
\end{equation*}
The previously measured $^{60}$Fe/Fe ratio in the Pacific Ocean sediment around 1.9\,Myr of about 3\,$\times$10$^{-16}$ can be compared to a $^{60}$Fe/Fe ratio in the Indian Ocean sediment of about 1.7\,$\times$10$^{-15}$, i.e.\ a factor of 0.18. Considering the sediment characteristics (Table \ref{tab:Sedimentdata}), however, the Pacific Ocean sediment should have a factor of 4.62 higher isotopic ratio than the Indian Ocean sediment. This gives rise to the discrepancy of 1\,--\,2 orders of magnitude with either an incomplete extraction of $^{60}$Fe or a global inhomogeneity being the cause.\\\\
\newcolumntype{Y}{>{\centering\arraybackslash}X}
\begin{table*}
\centering
\begin{tabularx}{\textwidth}{lYYYYY}  
\hline    \hline \\[-2ex]   
\textbf{Sediment} & \textbf{Rate} & \textbf{Extracted Fe} &  \textbf{Density} &  \textbf{Calc. ratio} &  \textbf{Meas. ratio}   \\
 & cm/kyr &  mg/g & g/cm$^{3}$ & & \\

\hline \\[-2.5ex]
Indian  \cite{Wallner2016}, auth. & 0.3   & 2.4 & 1.16 & 1 & 1\\
Pacific \cite{Ludwig2016}, CBD  & 1.9   & 0.14 & 0.68 & 4.62 & 0.18 \\
Pacific, this work  & 1.9  & 0.8 & 0.68 & 0.81 & $<$\,0.07\\
\hline    \hline
\end{tabularx}
\caption{Calculated vs.\ measured $^{60}$Fe/Fe ratios of the Pacific sediment normalized to the Indian Ocean sediment for two different chemical leachings. The sedimentation rate, extracted Fe concentration and density of the sediments are given as sediment characteristics. The measured ratio is a factor of 26\,--\,27 lower than the calculated ratio based on the Indian Ocean sediment and a homogeneous deposition of interstellar $^{60}$Fe on Earth. The deposition of $^{60}$Fe into geological archives is, therefore, clearly inhomogeneous on a global scale.}
\label{tab:Sedimentdata}
\end{table*}
The adopted leaching of the Pacific sediment in this work should yield a factor of 0.81 for the $^{60}$Fe/Fe ratio between the Pacific and the Indian Ocean sediment for a globally homogeneous deposition. This reduction is due to the stronger leaching and, therefore, dilution by terrestrial Fe from the sample. If $^{60}$Fe was insufficiently extracted before, however, the measured $^{60}$Fe/Fe ratio would increase from about 3\,$\times$10$^{-16}$ to about 1\,$\times$10$^{-15}$. The $^{60}$Fe/Fe ratio was measured at ANU by AMS, see \cite{Wallner2023, KollPhD} for details on $^{60}$Fe AMS measurements. The background corrected $^{60}$Fe/Fe ratio of (5.6\,$\pm$\,5.6)\,$\times$10$^{-17}$, insignificantly different from no interstellar $^{60}$Fe being detected, is at least one order of magnitude lower than in the Indian Ocean sediment. Considering now the expected factor of 0.81 between the two sediments and a measured factor of $<$\,0.07, the sediments are still discrepant by more than one order of magnitude. The uncertainties of the sediment characteristics on the percent level are negligible compared to the counting statistics of the measurement. This represents experimental evidence that the $^{60}$Fe deposition is globally inhomogeneous. The extent to which the deposition is inhomogeneous cannot be quantified yet to the required precision due to the low statistics and the variations between individual samples. Nevertheless, this is a promising starting point for future investigations. 
In light of the findings in this work, the internal discrepancy in fluence by a factor of 1.9 between the two sediments of Ludwig et al.\ \cite{Ludwig2016}, can now be interpreted as a consequence of global inhomogeneity resulting from their latitudinal distance of approximately 640\,km. This is in accordance with differing ocean conditions observed between the two sediments, as evidenced by their distinct sedimentation rates (6.1\,m/Myr and 19.3\,m/Myr) \cite{Ludwig2016}.
Considering the measured lower variability of $^{60}$Fe depositions in ferromanganese crusts compared to sediments \cite{KollPhD}, ferromanganese crusts might become the preferred archive to investigate the terrestrial fluence of $^{60}$Fe when uptake factors are constrained by lunar $^{60}$Fe inventories. Hydrogenetic ferromanganese crusts form from dissolved minerals and metals in sea-water which might render them less susceptible to local differences in scavenging efficiencies, bio-perturbation or diagenetic effects, see \cite{Christl2010} for a discussion of various sedimentary processes responsible for differences in radionuclide concentrations. Sediments with their higher sedimentation rates are valuable to constrain the timing and the terrestrial distribution of $^{60}$Fe influxes. Measurements of several sediments from different latitudes as well as ferromanganese crusts would constrain the global inhomogeneity as well as incorporation efficiencies of ferromanganese crusts.\\
It has to be concluded that the estimations of interstellar radionuclide influxes into geological archives require a location-dependent inhomogeneity factor $\kappa$ to account for the inhomogeneous deposition in addition to the already established incorporation efficiency $\epsilon_U$ for ferromanganese crusts. This assumes that sediments eventually incorporate $^{60}$Fe from seawater, while ferromanganese crusts incorporate only a fraction of the total $^{60}$Fe inventory. The location-dependent inhomogeneity factor applies to both archives. A relative inhomogeneity factor $\kappa_S$, relative to the Indian Ocean sediment of Wallner et al.\ \cite{Wallner2016}, has already been introduced \cite{KollPhD}.

\vspace{1cm}

\section{Conclusion}\label{conclusion}
Over the past 25 years, numerous geological archives, timescales and radionuclides were investigated to shed light on the deposition of interstellar dust on Earth. The investigated archives comprise deep-ocean crusts \cite{Knie1999a, Knie2004, Fitoussi2008, Wallner2016, Wallner2021, KollPhD}, nodules \cite{Wallner2016}, deep-sea sediments \cite{Fitoussi2008, Wallner2016, Ludwig2016, Wallner2020}, Antarctic snow \cite{Koll2019a, KollMSc} and lunar regolith \cite{Fimiani2016}. This work summarises our current projects to answer open questions about the origin of the recent $^{60}$Fe influx found in Antarctic snow \cite{Koll2019a} and sediments \cite{Wallner2020}, the exact timing of both $^{60}$Fe peaks and the influx of both $^{60}$Fe and $^{244}$Pu onto Earth and the Moon. The chemical extraction of minuscule amounts of live interstellar radionuclides from environmental samples is in some cases already well-developed but still requires further developments for radionuclides such as $^{182}$Hf. The first experimental evidence for an inhomogeneous influx of interstellar radionuclides is the starting point for future investigations of depositional patterns on Earth. The puzzle regarding the low $^{60}$Fe influx into the sediments of Ludwig et al. \cite{Ludwig2016} is now resolved by the inhomogeneous deposition of $^{60}$Fe into geological archives. The possibility that primary GCR radionuclides contribute to the radionuclide inventory in geological archives is intriguing and will be investigated further in the future.

\vspace{1cm}

\section*{Acknowledgements}\label{Ackn}
DK was supported by an AINSE Ltd. Postgraduate Research Award (PGRA).
This work was supported in part by the European Union (ChETEC-INFRA, project no. 101008324) and the Australian Research Council’s Discovery scheme, project numbers
DP180100495 and DP180100496.
The support for the Heavy Ion Accelerator Facility at ANU by NCRIS is highly appreciated.
DK wants to thank Thomas Faestermann for fruitful discussions about heavy ion fragmentation, Jenny Feige for the leaching of the Pacific sediment sample and Ulf Linnemann for providing deep-ocean nodules to develop chemical procedures. SM wants to thank Michael Kern and Laurenz Widermann for Hf-W-discussions and Johannes Sterba for INAA measurements.\\
The authors are grateful to the countless people working on this exciting topic over more than 25 years, from both the theoretical and experimental sides, in particular acknowledging the fundamental contributions from Thomas Faestermann and Gunther Korschinek.


\end{document}